\documentstyle[12pt]{article}

\begin{document}


\hsize=6.15in
\vsize=8.2in
\hoffset=-0.42in
\voffset=-0.3435in

\normalbaselineskip=24pt\normalbaselines

\begin{center}
{\large \bf Towards comparative theoretical neuroanatomy of the
cerebral cortex}
\end{center}

\vspace{0.15cm}

\begin{center}
{Jan Karbowski}
\end{center}

\vspace{0.05cm}

\begin{center}
{\it Sloan-Swartz Center for Theoretical Neurobiology,
Division of Biology 216-76, \\
California Institute of Technology,
Pasadena, CA 91125, USA \/}
\end{center}


\vspace{0.1cm}

\begin{abstract}
Despite differences in brain sizes and cognitive niches among mammals,
their cerebral cortices posses many common features and regularities. 
These regularities have been a subject of experimental investigation 
in neuroanatomy for the last 100 years. It is believed
that such studies may provide clues about cortical design principles
and perhaps function.
However, on a theoretical side there has been little interest, 
until recently, in studying quantitatively these regularities. 
This article reviews some attempts in this direction with an emphasis
on neuronal connectivity. It is suggested that the brain development
is influenced by different, conflicting in outcome, functional/biochemical 
constraints. Because of these conflicting constraints, it is hypothesized
that the architecture of the cerebral cortex is shaped by some global 
optimization plan.
\end{abstract}





\vspace{0.2cm}

\noindent To be published in {\it Journal of Integrative Neuroscience \/};
special issue on ``Neuromorphic models''.

\vspace{0.2cm}

\noindent Email: jkarb@cns.caltech.edu, jkarb@its.caltech.edu

\noindent Phone: (626)-395-5840, Fax: (626)-795-2397.

\vspace{0.3cm}

Running title: ``Theoretical neuroanatomy of the cortex''

\newpage

\section{Introduction}

It is generally believed that brains are evolutionary designed in such a way 
as to perform some functional computation which is vital for animals 
survival and reproduction, and possibly for higher cognitive functions 
[2]. In general, bigger mammals
have bigger brains - the brain volume $V_{b}$ scales with body volume
$V_{body}$ with the exponent around $3/4$ [2,26]. The origin of
this scaling law with this particular exponent is unknown, although there
were some suggestions that it may be a result of an increased number of 
sensory receptors on a surface body in bigger animals that
require more space for representation in the brain (e.g. [31]).
Also, motor output and homeostasis of the whole body, both of each are
controlled by the brain, may require more brain representation
in larger animals.
Despite big span in size (e.g. the volume of the mouse brain is 
$10^{4}$ times
smaller than that of the elephant), inter- and intraspecies variability,
brains of different species share common structures and many parameters 
associated with them exhibit striking regularities [7,9,16,19,23,28,36]. 
This similarity in structure and form may be an
indication of the same basic genetic design principles [31] governing
developmental processes. This article focuses on the regularities present
in the cerebral cortex, the part of the brain responsible for processing
of a sensory information and for higher cognitive function. In particular,
I shall review some recent theoretical approaches aiming at providing 
quantitative
framework for understanding neuroanatomical connectivity of the cortex.
It is believed that such approaches may ultimately turn out to be
helpful in deciphering cortical design principles and additionally may
be useful in providing link between ``structure and function'' [29].

The central theme of this review is the notion 
that architecture and size of the cerebral cortex are shaped by 
different constraints with conflicting outcomes. Some of these constraints
are related to maintaining functionality and some are connected to
biochemical/metabolic costs associated with cortical computations.
The hypothesis presented in this review is that, despite those competing
constraints, evolution has found ways to develop functional brains,
which represent a balanced design that is in some sense optimal. I shall
discuss both experimental data and recent theoretical approaches that
seem to point in this direction. In particular, I shall present the
current state of the art of the neuroanatomical data, and discuss what
still remains a challenge both experimental and theoretical.

\section{Invariants in the cerebral cortex}

Many scaling relations between cortical parameters are a direct consequence
of cortical invariants. There are several parameters which are roughly 
invariant with respect to brain size, across different cortical regions and
different species. These invariant parameters are:
(i) synaptic density $\rho_{s}$ [17,43],
(ii) surface density of neurons $\sigma_{n}$ [41],
(iii) the ratio of the number of excitatory to inhibitory synapses [19]
 (iv) cortical module size [34],
(v) density of short-range (intra-cortical) axons and dendrites [9].

From the above invariants one can derive interdependence relations 
between different cortical parameters [32,33]. To achieve
this, first let us introduce some definitions. Synaptic density is
defined as $\rho_{s}= N M/V_{g}$, where $N$ is the number of neurons in 
gray matter, $M$ is the average number of synapses per neuron, 
and $V_{g}$ is
the gray matter volume. Surface density of neurons is defined as 
$\sigma_{n}= N/W$, where $W$ is the total surface area of the cerebral 
cortex. Density of short-range axons is defined as $N L_{s}/V_{g}$, 
where $L_{s}$ is the average length of short-range (intra-cortical)
axons per neuron. 

If we use a scaling relationship between the total cortical surface area
$W$ and the gray matter volume $V_{g}$, $W\sim V_{g}^{0.9}$ [27], 
which is valid for large convoluted brains, then we obtain from the above
invariants and definitions that $M\sim V_{g}^{0.1}$ and roughly 
$M\sim N^{0.1}$. This means that
the number of synapses per neuron increases very weakly with brain size,
and additionally, that this number increases similarly weakly with the 
total number of neurons. The latter implies that cortical
networks become more and more sparse in terms of interconnectedness
as they get bigger, since the average connectivity
$M/N \sim N^{-0.9}$, i.e. it decays quickly with the number of neurons
(or the brain size). 
Why the number of synapses per neuron should increase with brain size? 
This may be a by-product of an expectation that the average
axon (dendrite) length per neuron should increase with brain size. 
The rationale for this is that axons should catch up, at least partially, 
with increased brain size in order to maintain some level of 
cross-communication with
other neurons in the network. Indeed, if we combine invariants 
(i) and (v), we obtain that the ratio $M/L_{s}$ is
brain size independent. Two conclusions follow from this relation.
First, an average inter-synapse distance is roughly constant across species. 
Second, the axon length $L_{s}$
scales with brain size in the same manner as the number of synapses does,
i.e. the axon length increases with the brain volume very slowly with the 
exponent 0.1. The uniformity of the inter-synapse interval distribution
maybe in some sense optimal for information processing and this may
cause the number of synapses per neuron to increase weakly with brain size.

Another consequence of the above invariants is that the number of neurons
contained in a module is brain size independent and this follows from
combining invariants (ii) and (iv). Since a cortical module is considered
to be an elementary unit processing information, this result may suggest 
that, in a first
approximation, brains of different sizes use essentially the same local
computational mechanisms. The differences between brains functioning may arise 
from a larger-scale organization, i.e. connectivity patterns between
modules and cortical areas.  

The fact that the number of excitatory to inhibitory synapses is constant
across different species with different brain volumes may suggest that
there exist mechanisms in the brain that
try to maintain a global balance between excitation and inhibition 
[44,52]. Such a balance can be achieved by homeostatic processes
that can dynamically adjust the number of
synapses and their strength [50]. From a functional
point of view, the balance between excitation and inhibition
is necessary for a permanent regulation of neuronal 
activity. In fact, it is an efficient way of preventing both
disastrous hyperactivation (when excitation dominates) and equally
disastrous, for the brain function, inactivation 
(when inhibition dominates).  

Why are there invariants in the cerebral cortex at all? The precise answer
to this question is unknown, however the very fact of their existence
can hint us about possible mechanisms that shape architecture of
the cortex. Recently, an interesting theoretical idea was proposed
[15] that can be used to address that question.
These authors considered ``thought experiments'' with perturbing
some cortical processes and looked how these perturbations influenced
local axonal conduction delays in the cortex. They found that conduction
delays are minimal when both the volume of axons and the volume of 
synapses constitute $3/5$ of the total cortical volume of gray matter. 
That prediction is consistent with experimental data for axonal and 
dendritic volumes [15]. Since the derivation of 
this fraction is quite general and brain size independent, it is possible 
that minimization of conduction delays is the main factor behind some of 
the cortical invariants (i.e. invariants (i) and (v)).

\section{Local vs. large-scale connectivity}

The cerebral cortex is organized differently at different levels.
On a microscopic scale, neurons are connected in sparse local circuits
[8,21] with a probability of a direct connection decreasing rapidly with a
distance [24]. The average probability of a contact between
two neurons, defined as the ratio of the average number of synapses per
neuron to the total number of neurons, can be computed from the above 
invariants [32,33]. It is given by [34]

\begin{equation}
p\sim V_{g}/N^{2}, 
\end{equation}\\
and it decays with brain size as $V_{g}^{-0.8}$ [34,49]. 

Early studies [9,25]
suggested that local wiring pattern is stochastic, that is, neurons tend to 
connect with other neurons in a random fashion. Such conclusion was
motivated by a discovery that synaptic interbouton intervals along
an axon in the rodent cortex are distributed according to a Poisson process 
and there was no correlation between them [25]. However recent 
body of evidence suggests that neurons are highly selective in choosing their
targets (e.g. [11,54]). Only certain classes of
neurons are connected by a given class, and a such defined connectivity
pattern seems to be almost deterministic. Thus, there seems to be
little correlation between apparent stochasticity in the bouton distribution 
and selectivity in neuronal connections. 

More recently Anderson et al [4] studied the distribution of interbouton
intervals in more detail and across different neuronal classes.
These authors found in the
cat visual cortex that interbouton intervals in initial axonal segments are 
distributed according to a Poisson process, but in most other segments 
and cases they can be fitted well to a gamma distribution except for 
very long intervals. At those long intervals distributions exhibit
heavy tails, however they could not be fitted to a single power law.
Additionally, they found that parameters characterizing each distribution 
are very similar for cells within the same class but differ among classes. 
These results indicate that the synapses are distributed in somewhat more 
ordered way than was thought previously, and these findings are consistent 
with the idea of specificity. It remains a challenge for the future, both 
experimental and theoretical, to develop models of such connectional
specificity (see also Conclusions section).

One can draw also another conclusion from the interbouton interval
distribution. This distribution should correlate to a certain degree 
with the distribution of
the number of synapses per neuron, since for a given axonal length
the number of synapses is inversely proportional to the average interbouton
interval. The presence of heavy tails in the distribution of interbouton 
intervals might translate to the distribution of synapses per neuron 
having such heavy tails as well. However, one should be cautious with
this, since there is no simple mathematical one-to-one relationship
between these two quantities.
This feature of long tails in the number of synapse distribution, if proved 
experimentally correct, can have interesting implications for cortical 
computation (see e.g. [48]).  

On a macroscopic scale cortical networks are organized into areas with
distinct cytoarchitectonic and neuroanatomical properties. 
Large-scale connectivity between areas has been investigated by Young
and colleagues [42,55,56] in cat and monkey. In a series of papers they
classified the area connectivity using multidimensional scaling
method [56]. The main conclusion from their work is that 
areas tend to
connect mostly with their neighbours and only rarely with remote areas.
This architecture resembles the so-called ``small world'' networks
[53], which seem to possess
a suitable structure for efficient communication between network
components [47]. There are two main classes of small 
world networks that have
become an object of intense current research. One class is known 
as Erdos-Renyi networks [22], in which the distribution of the number 
of connections has
a pronounced peak at some finite value that can be approximated by
a gamma distribution. Second class
of small world networks that has received an extraordinary 
attention recently, is known as scale-free networks [6]. In this
type of networks the distribution of connections has a long tail and
follows a power law. It is interesting to investigate which of these
two types is actually realized in the cortical large-scale organization.
In Figs. 1 and 2 we plot the cumulative distribution of connections
between cortical areas for cat (Fig. 1) and for monkey (Fig. 2)
using data of connectivity matrices from Young et al [54].
The cumulative distribution $C(k)$ is defined as a proportion of 
the number of areas having at least $k$ connections with other areas. 
It would give a power law if a regular distribution, defined as 
a proportion of the number of areas
having precisely $k$ connections, had a power law decay. 
The log-log plots do not yield straight lines in either of these cases,
which indicates that cortico-cortical connectivity is not organized
as a scale-free network. This result taken together with the possibility
of long-range tails in the distribution of the number of synapses
suggest that microscopic and macroscopic cortical organization can 
differ not only quantitatively but also qualitatively.

What are the factors that influence the connectivity between cortical
areas? This question was a subject of a theoretical approach which
aimed at relating connectivity to other cortical parameters [34].
The average connectivity $Q$ between two arbitrary chosen areas A and B, 
(this is a different quantity than the average probability of connection 
$p$ between two neurons) is defined as a probability that
at least one of the cortical modules (columns or barrels) in A
is connected to B. It was found that $Q$ depends on other parameters 
in the following from [34]:

\begin{equation}
Q\approx 1 - \exp\left(- \frac{a L_{0}^{2}}{\xi^{2}K^{2}}\right),
\end{equation}\\
where $L_{0}$ is the average axon length in white matter, $\xi$ is
the average linear size of a cortical module, $K$ is the total number 
of cortical
areas, and the dimensionless parameter $a$ characterizes a particular
cortical geometry and a pattern of axonal organization in white matter.
From this formula, it follows that the connectivity depends mainly
on two factors: the average axon length in white matter and the number
of cortical areas. 

Equation (2) is important for two reasons. First, it can have a practical
application in determining axonal length in white matter, since it is 
difficult to do it directly experimentally. Second, it was found, 
based on scaling laws for the above parameters, that the average connectivity 
$Q$ is either only weakly dependent or independent of brain size [34]. 
This is in contrast to the connectivity $p$ between
neurons (see Eq. (1)), which decays quickly with brain size.
The finding that $Q$ only weakly decays with the brain volume, also
provides some hint about the large-scale cortical connectivity. It
may suggest that brain evolutionary design tries to prevent isolation of 
cortical areas as the brain gets bigger.

\section{Optimal wiring}

Considerations of the previous section suggest that the wiring
pattern in the cortex is not random but there is some plan associated
with it. This is not a new idea - it has a long history in neuroscience
dating back to early neuroanatomists like Cajal [10].
But what is precisely that wiring plan in the cortex? 
Is it the same in gray matter as in white matter? These are not 
easy questions to answer in full detail because of the complexity of
different neuronal types and thousands of connections between them in gray
matter on the one hand, and technical problems with investigating axonal
organization in white matter, on the other hand.
Despite these difficulties, there are some reasons to believe
that the connectivity pattern in the cortex is somehow optimized. 
One strongly advocated optimization 
principle related to gray matter
is called the principle of minimal axon length [14,15,39,51],
and states that
the total axonal length or equivalently the axon length per neuron
(if we divide the total axonal length by the total number of neurons)
should be as small as possible in order for the cortex to be functional. 
A support for this hypothesis was provided by Cherniak [13]
by analyzing data from
the nervous system of a nematode worm
{\it Caenorhabditis elegans, \/} the only organism fully characterized
in terms of connectivity. He has found
that the total length of neural connections is indeed 
minimized. From a theoretical point of view, the demand of minimal
axon length is related primarily to the demand of small conduction
delays between neurons [15]. Large delays would interfere
with efficient information exchange between neurons and this could lead
to loss of some functions, which is clearly undesirable. Thus, one can
associate optimal wiring in the cortical gray matter with the requirement
of minimal conduction delays, which is equivalent to the principle of
minimal axon length (although the axonal length per neuron is not exactly
the same as the maximal axonal pathlength, which is more directly related
to conduction delays, these two quantities should be strongly correlated).  

All the above considerations were related to the wiring pattern on a
level of local circuits in the gray matter. It is interesting to ask
if the same principle applies to long-range (cortico-cortical) connections
via white matter? Recently, this question was addressed [35] 
in the macaque brain. It was found that 11 cortical areas
in the prefrontal cortex are indeed connected through the axons that
minimize their total length. The calculation was based on all possible
arrangements of the cortical areas and it was
found that their actual positioning
in the brain is the one that minimizes the wire length.  
Although, this computation was performed only in a limited part of the 
cortex, there is a belief that it can be generalized to the 
whole cortex, thus providing yet another
support for the principle of minimal axon length/conduction delays.

Is wiring length the only quantity that is evolutionary optimized in the
brain? Other candidates for optimization on a large scale can include: 
metabolic energy consumed by the whole cortex, the number of cortical areas,
and some abstract complexity. That processes operating in the brain 
try to minimize their 
metabolic expenditure should not be surprising if one
recalls that the brain is energetically an expensive tissue, a hypothesis
put forward by Aiello and Wheeler [1]. The metabolic energy rate
of the whole cortex at rest scales with the gray matter volume as 
$V_{g}^{0.8}$ [26,34], which implies that energy consumption
per 1 g of the gray matter decreases with brain size. Using this experimental
result and another fact that glutamatergic excitatory synapses are the
main users of metabolic energy [5,45,46], one
can derive that the number of active synapses at any given instant should
decay with brain size as $V_{g}^{-0.2}$ [34]. This result is
consistent with the notion that brains may minimize their metabolism 
as well [3,38].

The increase in the number of cortical areas with brain size has been 
advocated by Kaas [30,31]. By having more areas, brains can perform
more functional tasks in local specialized circuits, thus restricting 
activity to specific regions. This can be more optimal in
terms of saving biochemical resources than could be more distributed 
large-scale processing.

Recently it has been suggested by Sporns et al [47] that the large-scale
cortical organization of ``small world'' type can support highly
complex dynamics of neuronal activity. Similar type of dynamics
has been observed {\it in vivo \/} [52] and this led these authors to
propose that the cortical architecture optimizes some abstractly
defined complexity.

In the next section, I shall introduce 3 hypothetical functional principles
of the brain operation that constitute brain's ``computational power''.
I shall argue that the brain architecture cannot optimize all
quantities associated with these principles simultaneously. Rather, 
the optimal design is a compromise
between optimizing each of those quantities separately.

\section{Trade-offs in the brain design}

The first observation that the minimization of axonal length itself
cannot be ``the best solution'' to the brain design was provided
in [32,33]. The argument is that, on average, if axonal
length is small then more synaptic steps are needed to connect
two arbitrary neurons in the network, i.e. communication in the network 
is less efficient. 
Since synapses consume a large portion of metabolic energy [45,46], 
it implies that decreasing axonal length causes larger metabolic
use if no function is supposed to be lost. However, the brain has
a limited amount of energy available that is controlled by body biochemistry
and this leads to a trade-off. Thus the brain design must choose 
a compromise between 
the two extremal solutions, and it is impossible to have both short axons
and low metabolic energy rate at the same time. This reasoning can be put 
in a more quantitative language that takes into account the cortical 
invariants and architecture [32].

Another argument indicating that brains are under pressure of different
sorts of constraints was presented recently [34]. It was
shown that bigger brains could face size and architectonic problems, if
some functional requirements were satisfied simultaneously. 
Three simple hypothetical functional principles were proposed (for
extended discussion, see [34]): (i) the number of areas should increase
with brain size as quickly as possible, (ii) the area-area connectedness
should not decay with brain size, (iii) the temporal delays between
areas should not increase with brain size. Obviously, one can imagine
more similar ``reasonable'' principles operating both on a large and
local scales. However, for the sake of argument, let us focus on the
above three, characterizing large-scale cortico-cortical organization
of the cortex. I assume that these 3 functional principles constitute
the brain's computational power.

If we assume that white matter is composed primarily of
cortico-cortical axons, then one can derive [34] that the ratio
of white matter to gray matter volumes $V_{w}/V_{g}$ obeys

\begin{equation}
V_{w}/V_{g} \sim V_{g}^{-0.1} \frac{K^{3} Q^{3/2}}{\tau^{2}},
\end{equation}\\
where $\tau$ is the average conduction delay between cortical areas.
From this, it follows that if evolution wants to keep this delay relatively
brain size independent and simultaneously to increase the number of
areas with brain size at high rate, then this would lead to an 
excessive scaling
of the white matter volume with gray matter (longer long-range axons). 
This, in turn, would imply
bigger brains as a whole, and that would cause mechanical problems
for a body. To prevent this type of design catastrophe, evolution has
to compromise part of the brain's theoretical computational power. 
It is done, as experimental data shows, by
simultaneously: (i) increasing slightly temporal delays as brains get 
bigger [40], (ii) decreasing the rate of growth of the number of areas with
brain size [12], and (iii) decreasing slightly connectivity $Q$ with
brain size [34].

It is very likely that brains have to face more functional/architectonic
compromises that wait to be discovered. These may involve constraints
on cellular and molecular levels (e.g. [37]). 
For instance, different dendritic shapes can be a result of such 
compromises. Also, axons may be under mechanical stress which may 
lead effectively to cortical convolutions [51], which in turn can
reduce significantly the total axonal length [51].

\section{Conclusions}

In this review, there have been presented different types of constraints
which may affect the developmental design of the cerebral cortex.
Because many of these constraints lead to conflicting outcomes, it
is suggested that there exists some global optimal design plan that
guides the brain throughout the development. Such an optimal design
is probably a product of an evolutionary pressure on genes, which
control development. Thus, in order to construct theories of optimal
cortical design, one has first to gain a much better understanding
of the influence of genes and gene products (proteins) on the developmental
process. This is challenging, however, both experimentally and theoretically.
From a theoretical perspective, it is not a trivial question to relate
a local nature of molecular/chemical interactions of gene products to
the globality of the organizational plan. It seems that diffusion may
play some role in this connection, since it enables transport of
chemicals over large distances, and thus leads effectively (although
with some delay that can vary between a fraction of a second to days) 
to global communication [18].

The examples shown in this review may suggest that the evolutionary brain
design had to optimize not one parameter but probably many parameters 
simultaneously in order to make brains functional. Such optimization is
not a trivial problem and it leads to multiple trade-offs in some
abstract multidimensional parameter space. It is possible that
brain design solves this multidimensional optimization problem by
adjusting different parameters in order to operate in a ``global minimum''.
The challenge for the future is to try to identify the relevant
optimization parameters, and to verify this hypothesis.
However, to achieve this, more reliable neuroanatomical data 
across many species are needed. For instance, only for two mammals:
monkey and cat, we know the detailed large-scale connectivity matrix [56].
Quantitative local connectivity has been investigated to some rather 
modest extent only in mouse and rat [24,25]. This is clearly too little
for theoretical developments. More effort should be put in such studies
for other animals, as well.

\noindent{\bf Acknowledgments}

The work was supported by the Sloan-Swartz fellowship at Caltech.

\vspace{1.5cm}

\noindent {\bf References} 

\noindent [1] Aiello, L.C., and Wheeler, P. The expensive-tissue hypothesis:
The brain and digestive system in primate evolution. {\it Current
Anthropology, \/} {\bf 36}, (1995), pp. 199-221. 

\noindent [2] Allman, J.M. {\it Evolving Brains. \/} (Freeman, New
York, 1999).

\noindent [3] Ames III, A. CNS energy metabolism as related to function.
{\it Brain Research Reviews \/} {\bf 34}, (2000), pp. 42-68.

\noindent [4] Anderson, J.C., Binzegger, T., Douglas, R.J., and Martin, 
K.A.C. Chance or design? Some specific considerations concerning
synaptic boutons in cat visual cortex. {\it J. Neurocytology \/} {\bf 31},
(2002), pp. 211-229.

\noindent [5] Attwell, D., and Laughlin, S.B. An energy budget for
signaling in the gray matter of the brain. {\it J. Cerebral Blood Flow
and Metabolism \/} {\bf 21}, (2001), pp. 1133-1145.

\noindent [6] Barabasi, A.L., and Albert, R. Emergence of scaling in
random networks. {\it Science \/} {\bf 286}, (1999) pp. 509-512.

\noindent [7] Barton, R.A., and Harvey, P.H. Mosaic evolution of brain
structure in mammals. {\it Nature \/} {\bf 405}, (2000), pp. 1055-1058.

\noindent [8] Braitenberg, V. Cell assemblies in the cerebral cortex.
In {\it Theoretical approaches to complex systems. \/} Eds. R. Heim and
G. Palm (Springer-Verlag, Berlin, 1978).

\noindent [9] Braitenberg, V., and Sch{\"u}z, A. {\it Cortex: Statistics
and Geometry of Neuronal Connectivity. \/} (Springer, Berlin, 1998).

\noindent [10] Cajal, S.R. {\it Histology of the Nervous System
of Man and Vertebrates. \/} (Oxford Univ. Press, New York, 1995), vol. 1.

\noindent [11] Callaway, E.M. Cell type specificity of local cortical
connections. {\it J. Neurocytology \/} {\bf 31}, (2002), pp. 231-237.

\noindent [12] Changizi, M.A. Principles underlying mammalian neocortical
scaling. {\it Biol. Cybern., \/} {\bf 84}, (2001), pp. 207-215.

\noindent [13] Cherniak, C. Component placement optimization in the
brain. {\it J. Neuroscience \/} {\bf 14}, (1994), pp. 2418-2427.

\noindent [14] Cherniak, C. Neural component placement.
{\it Trends Neurosci. \/} {\bf 18}, (1995), pp. 522-527.

\noindent [15] Chklovskii, D.B., Schikorski, T., and Stevens, C.F.
Wiring optimization in cortical circuits. {\it Neuron \/} {\bf 34},
(2002), pp. 341-347.

\noindent [16] Clark, D.A., Mitra, P.P., and Wang, S.S.H. Scalable
architecture in mammalian brains. {\it Nature \/} {\bf 411}, (2001),
pp. 189-193.

\noindent [17] Cragg, B.G. The density of synapses and neurones in the
motor and visual areas of the cerebral cortex. {\it J. Anatomy\/}
{\bf 101}, (1967), pp. 639-654.

\noindent [18] Crick, F. Diffusion in embryogenesis. {\it Nature \/}
{\bf 225}, (1970), pp. 420-422.

\noindent [19] DeFelipe, J., Alonso-Nanclares, L., and Avellano, J.
Microstructure of the neocortex: Comparative aspects. {\it J.
Neurocytology \/} {\bf 31}, (2002), pp. 299-316.

\noindent [20] De Winter, W., and Oxnard, C.E. Evolutionary radiations
and convergences in the structural organization of mammalian brains.
{\it Nature \/} {\bf 409}, (2001), pp. 710-714.

\noindent [21] Douglas, R.J., and Martin, K.A.C. Opening the grey box.
{\it Trends Neurosci., \/} {\bf 14}, (1991), pp. 286-293.

\noindent [22] Erdos, P., and Renyi, A. {\it Publ. Math. Inst. Hung.
Acad. Sci. \/} {\bf 5}, (1960), pp. 17-28.

\noindent [23] Finlay, B.L., and Darlington, R.B. Linked regularities
in the development and evolution of mammalian brains. {\it Science \/}
{\bf 268}, (1995), pp. 1578-1584.

\noindent [24] Hellwig, B. A quantitative analysis of the local
connectivity between pyramidal neurons in layers 2/3 of the rat
visual cortex. {\it Biol. Cybern., \/} {\bf 82}, (2000), pp. 111-121.

\noindent [25] Hellwig, B., Schuz, A., and Aerstsen, A. Synapses on
axon collaterals of pyramidal cells are spaced at random intervals:
A Golgi study in the mouse cerebral cortex. {\it Biol. Cybern., \/}
{\bf 71}, (1994), pp. 1-12.

\noindent [26] Hofman, M.A. Energy metabolism, brain size and longevity
in mammals. {\it Quarterly Review of Biology \/} {\bf 58}, (1983),
pp. 495-512.

\noindent [27] Hofman, M.A.  Size and shape of the cerebral cortex
in mammals. I. The cortical surface. {\it Brain Behav. Evol. \/} {\bf 27},
(1985), pp. 28-40.

\noindent [28] Hofman, M.A. On the evolution and geometry of the brain
in mammals. {\it Prog. Neurobiol. \/} {\bf 32}, (1989), pp. 137-158.

\noindent [29] Jerison, H.J. {\it Brain size and the evolution of mind. \/}
(Am. Mus. Natl. Hist., New York, 1991).

\noindent [30] Kaas, J.H. The evolution of isocortex. {\it Brain Behav.
Evol. \/} {\bf 46}, (1995), pp. 187-196. 

\noindent [31] Kaas, J.H.  Why is brain size so important: Design
problems and solutions as neocortex gets bigger or smaller. {\it Brain
Mind \/} {\bf 1}, (2000), pp. 7-23.

\noindent [32] Karbowski, J. Optimal wiring principle and plateaus
in the degree of separation for cortical neurons. 
{\it Physical Review Lett. \/} {\bf 86}, (2001), pp. 3674-3677. 

\noindent [33] Karbowski, J. Optimal wiring in the cortex and neuronal
degree of separation. {\it Neurocomputing \/} {\bf 44-46}, (2002),
pp. 875-879. 

\noindent [34] Karbowski, J. How does connectivity between cortical areas
depend on brain size? Implications for efficient computation.
{\it J. Comput. Neurosci. \/} {\bf 15}, (2003), pp. 347-356; 
arXiv:q-bio.NC/0310015 v1.

\noindent [35] Klyachko, V.A., and Stevens, C.F. Connectivity
optimization and the positioning of cortical areas. {\it Proc. Natl.
Acad. Sci. USA \/} {\bf 100}, (2003), pp. 7937-7941.

\noindent [36] Krubitzer, L. The organization of neocortex in mammals:
Are species differences really so different? {\it Trends Neurosci. \/}
{\bf 18}, (1995), pp. 408-417.

\noindent [37] Krubitzer, L., and Huffman, K.J. Arealization of the 
neocortex in mammals: genetic and epigenetic contributions to the 
phenotype. {\it Brain Behav. Evol. \/} {\bf 55}, (2000), pp. 322-335.

\noindent [38] Laughlin, S.B., de Ruyter van Steveninck, R.R., and Anderson,
J.C. The metabolic cost of neural information.
{\it Nature Neurosci. \/} {\bf 1}, (1998), pp. 36-41.

\noindent [39] Mitchison, G. Axonal trees and cortical architecture.
{\it Trends Neurosci. \/} {\bf 15}, (1992), pp. 122-126.

\noindent [40] Ringo, J.L., Doty, R.W., Demeter, S., and Simard, P.Y. 
Time is of the essence: A conjecture that hemispheric specialization arises
from interhemispheric conduction delay.
{\it Cereb. Cortex \/} {\bf 4}, (1994), pp. 331-343.

\noindent [41] Rockel, A.J., Hiorns, R.W., and Powell, T.P.S. 
The basic uniformity in structure of the neocortex. {\it  Brain \/} 
{\bf 103}, (1980), pp. 221-244.

\noindent [42] Scannell, J.W., Young, M.P., and Blakemore, C. Analysis
of connectivity in the cat cerebral cortex.
{\it J. Neurosci.\/} {\bf 15}, (1995), pp. 1463-1483.

\noindent [43] Sch{\"u}z, A., and Demianenko, G. Constancy and variability 
in cortical structure. A study on synapses and dendritic spines in hedgehog
and monkey. {\it J. Brain Res. \/} {\bf 36}, (1995), pp. 113-122.

\noindent [44] Shu, Y.S., Hasenstaub, A., and McCormick, D.A.
Turning on and off recurrent balanced cortical activity. {\it Nature \/}
{\bf 423}, (2003), pp. 288-293.

\noindent [45] Shulman, R.G., and Rothman, D.L. Interpreting functional
imaging studies in terms of neurotransmitter cycling. {\it Proc. Natl. Acad.
Sci. USA \/} {\bf 95}, (1998), pp. 11993-11998.

\noindent [46] Sibson, N.R., Dhankar, A., Mason, G.F., Rothman, 
D.L., Behar, K.L.,
and Shulman, R.G. Stoichiometric coupling of brain glucose metabolism
and glutamatergic neuronal activity. {\it Proc. Natl. Acad. Sci. USA \/}
{\bf 95}, (1998), pp. 316-321.

\noindent [47] Sporns, O., Tononi, G., and Edelman, G.M. Theoretical
Neuroanatomy: Relating anatomical and functional connectivity in graphs
and cortical connection matrices.
{\it Cereb. Cortex \/} {\bf 10}, (2000), pp. 127-141.

\noindent [48] Stauffer, D., Aharony, A., Costa, L.D., and Adler, J. 
Efficient Hopfield pattern recognition on a scale-free neural network.
{\it Eur. Phys. Journ. B \/} {\bf 32}, (2003), pp. 395-399.

\noindent [49] Stevens, C.F. How cortical interconnectedness varies
with network size. {\it Neural Comput. \/} {\bf 1}, (1989), pp. 473-479.

\noindent [50] Turrigiano, G.G., and Nelson, S.B. Homeostatic plasticity
in the developing nervous system. {\it Nature Rev. Neurosci. \/}
{\bf 5}, (2004), pp. 97-107.

\noindent [51] van Essen, D.C. A tension-based theory of morphogenesis
and compact wiring in the central nervous system. {\it Nature \/}
{\bf 385}, (1997), pp. 313-318.

\noindent [52] van Vreeswijk, C., and Sompolinsky, H. Chaos in neuronal
networks with balanced excitatory and inhibitory activity.
{\it Science \/} {\bf 274}, (1996), pp. 1724-1726.

\noindent [53] Watts, D.J., and Strogatz, S.H. Collective dynamics of
``small-world'' networks. {\it Nature \/} {\bf 393}, (1998), pp. 
440-442.

\noindent [54] White, E.L. Specificity of cortical synaptic connectivity;
emphasis on perspectives gained from quantitative electron microscopy.
{\it J. Neurocytology \/} {\bf 31}, (2002), pp. 195-202.

\noindent [55] Young, M.P. Objective analysis of the topological
organization of the primate cortical visual system.
{\it Nature \/} {\bf 358}, (1992), pp. 152-155.

\noindent [56] Young, M.P., Scannell, J.W., and Burns, G. 
{\it The analysis of cortical connectivity. \/} 
(Landes, Austin, TX, 1995).

\newpage

{\bf \large Figure Captions}

Fig. 1a\\
Cumulative distribution of the large-scale connectivity between
cortical areas for the cat cortex.

\vspace{0.3cm}

Fig. 1b\\
Log-Log plot of the cumulative distribution of the large-scale 
connectivity between cortical areas for the cat cortex.

\vspace{0.3cm}

Fig. 2a\\
Cumulative distribution of the large-scale connectivity between
cortical areas for the monkey cortex.

\vspace{0.3cm}

Fig. 2b\\
Log-Log plot of the cumulative distribution of the large-scale 
connectivity between cortical areas for the monkey cortex.

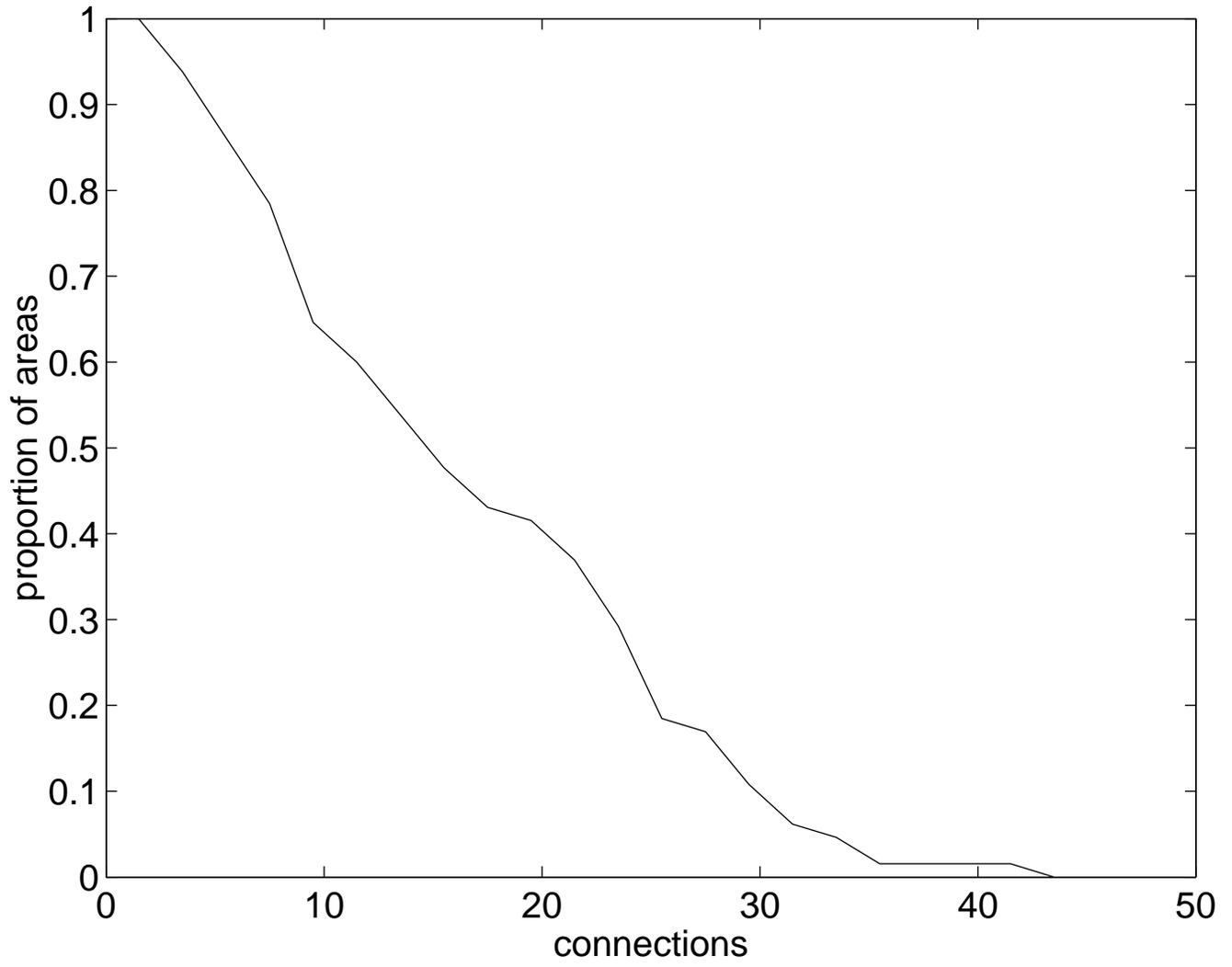
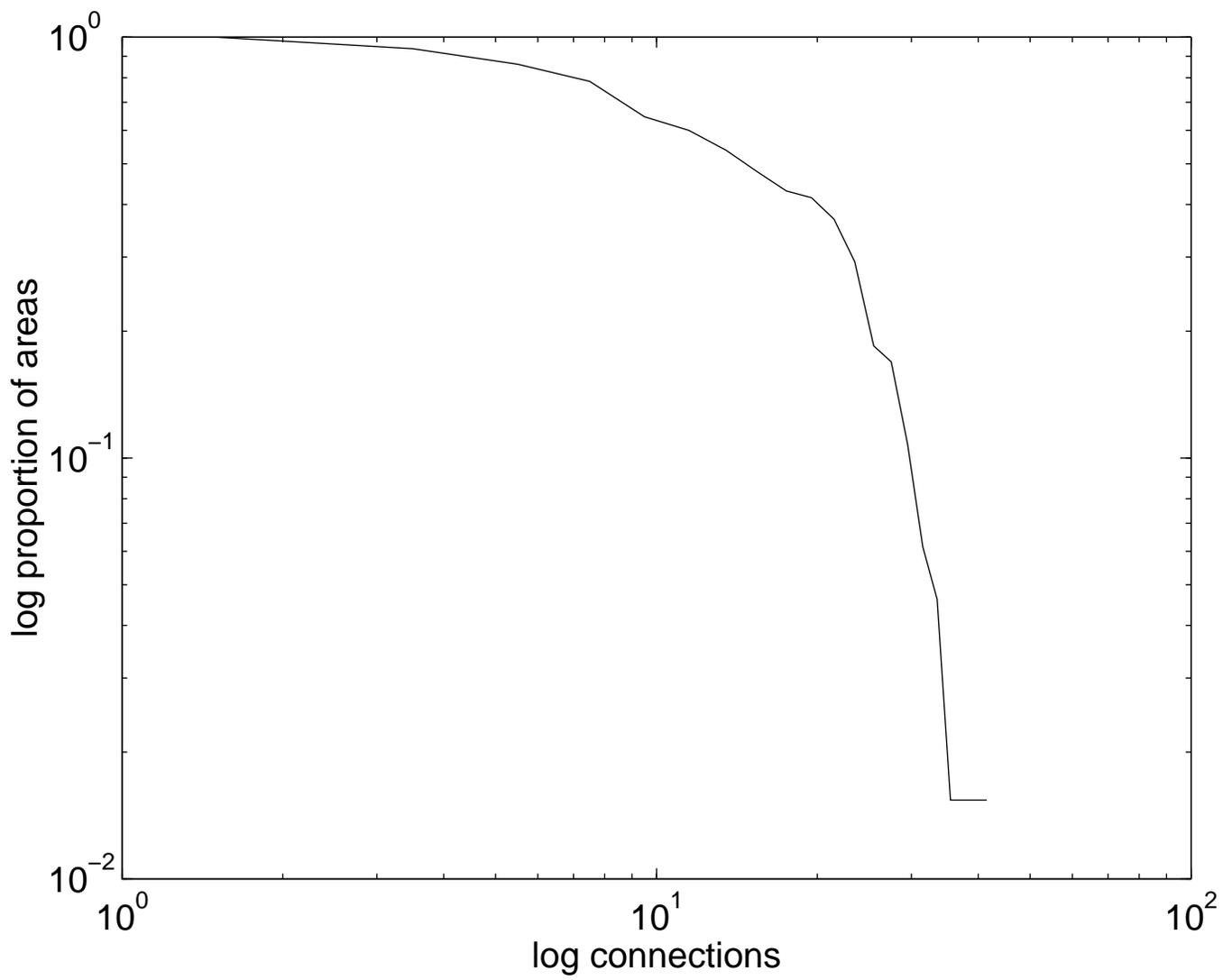
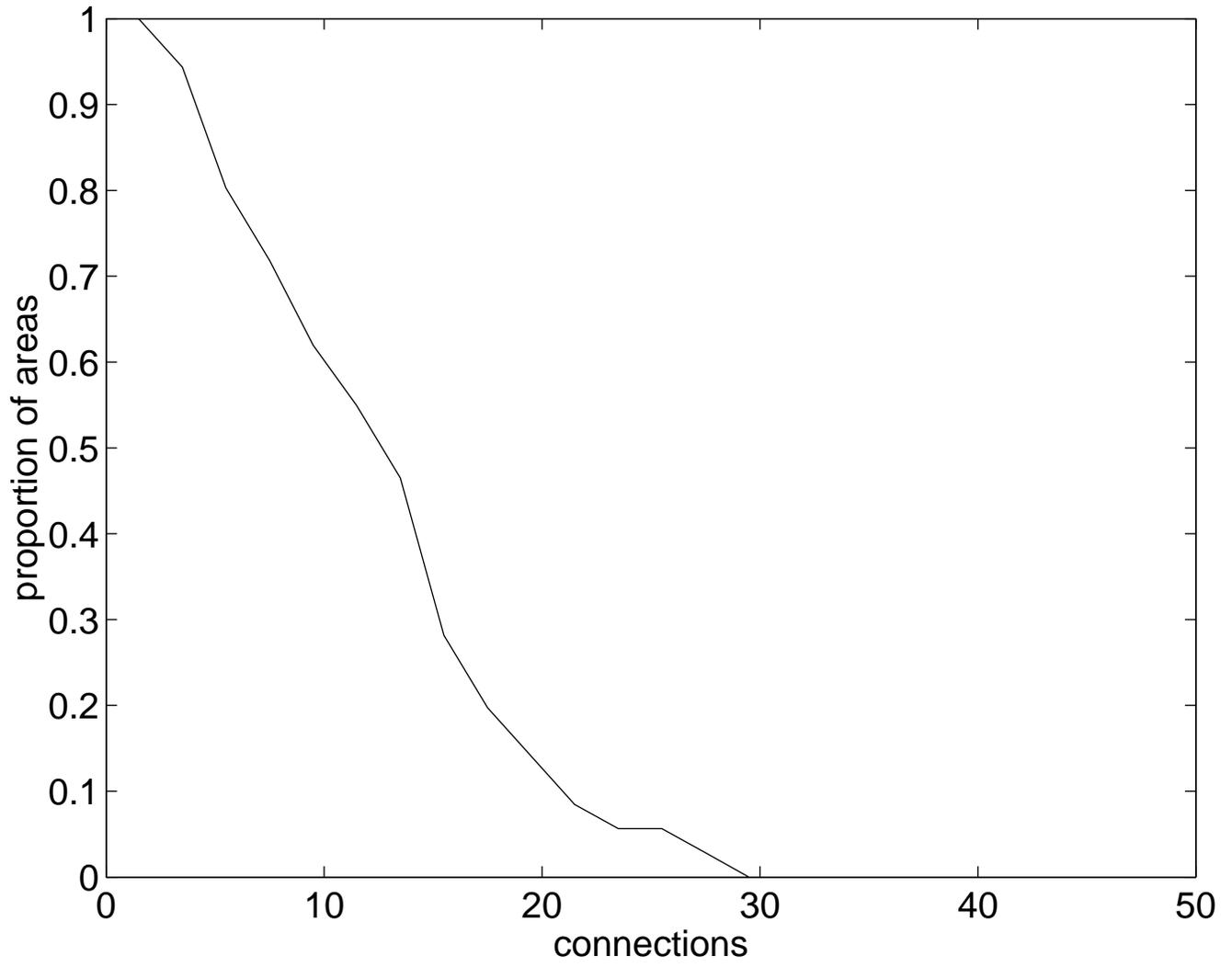
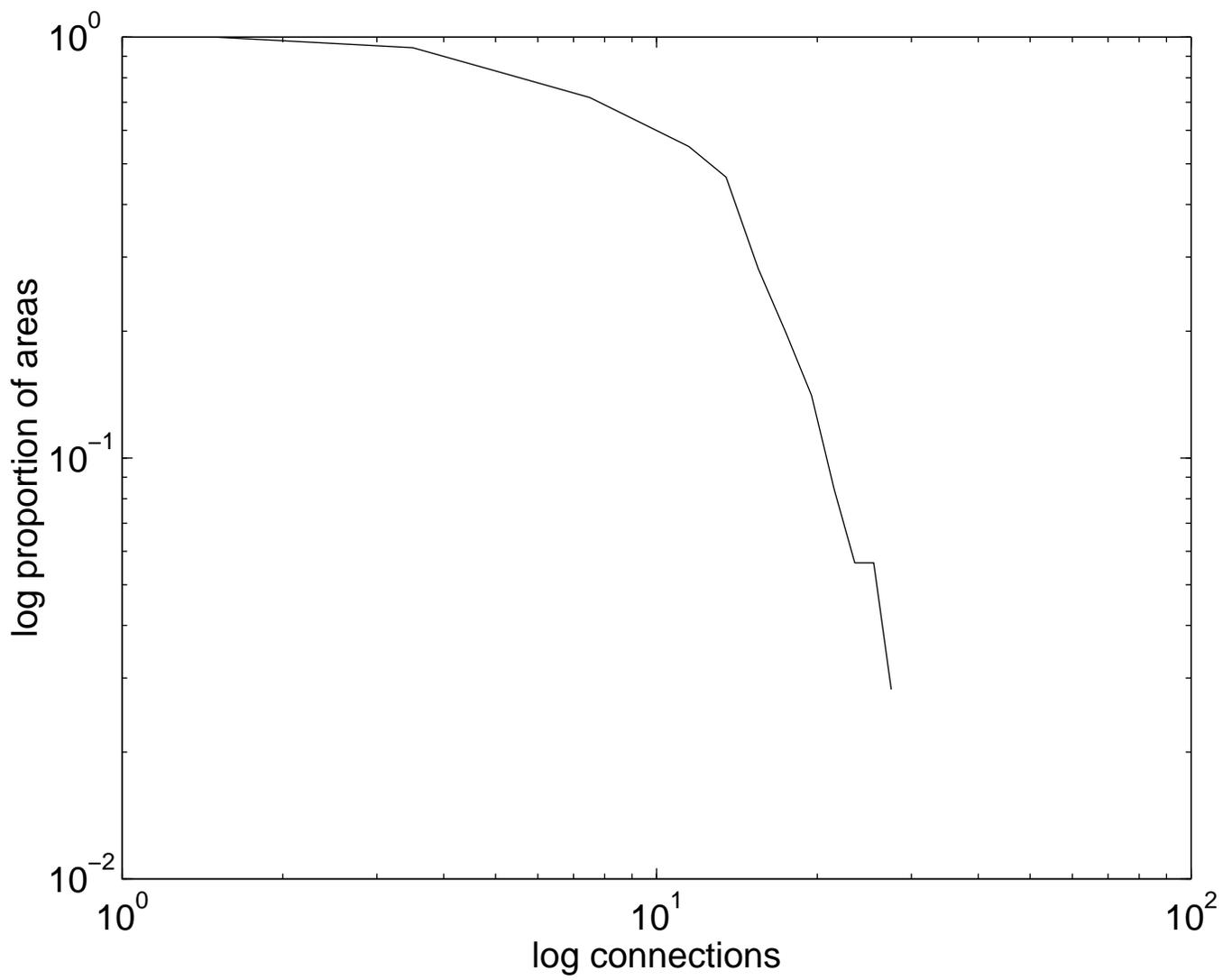
\end{document}